%% file: manuscript_b.tex
\documentclass[11pt]{article}
\usepackage[margin=1in]{geometry}
\usepackage{amsmath}
\usepackage{amssymb}
\usepackage{graphicx}
\usepackage{url}
\usepackage{xurl}
\usepackage{setspace}
\usepackage{tabularx}
\usepackage[skip=6pt]{caption}
\graphicspath{{figures/}}

\input{tables/numbers}
\input{tables/counts}

\newcommand{\archivedoi}{10.5281/zenodo.24261327}

\newif\ifanonymous
\ifdefined\paperbanonymous \anonymoustrue \else \anonymousfalse \fi
\ifanonymous
  \usepackage{lineno}
  \linenumbers
  \renewcommand{\archivedoi}{[removed for anonymous review]}
  \newcommand{\ethicsbody}{the author's institution's research ethics guidance}

  \newcommand{\referencelist}{references_b_anonymous}
\else
  \newcommand{\ethicsbody}{Trinity College Dublin's research ethics guidance}
  \newcommand{\referencelist}{references_b}
\fi

\title{New rules, old habits: across seven team sports, each league
keeps its own way of using the rules}
\ifanonymous
  \author{}
\else
  \author{Gustavo Pedro Ricou\\[2pt]
  \small School of Computer Science and Statistics, Trinity College Dublin,
  Dublin, Ireland\\
  \small \url{pedrorig@tcd.ie}}
\fi
\date{}

\newcommand{\sidewayswidth}{0.99\textheight}
\newcommand{\turnsideways}[1]{\rotatebox{90}{#1}}
\newcommand{\registerfont}{\fontsize{7.5}{8.5}\selectfont}
\begin{document}
\maketitle
\onehalfspacing

\begin{abstract}
Studies of a rule change usually report one average effect, for one
league or several combined. We asked whether a rule change reorders
leagues, in \pbAllLeagues{} leagues of seven sports. When the
substitution limit rose from three to five, the leagues already ranked
from most substitutions to fewest, as known for the biggest leagues. In
their first season under five, the \pbPreOrderN{} top leagues kept that
order (rank correlation \pbPreOrder; one means the same order, zero
unrelated orders). That is about as firmly as from one ordinary
season to the next. A prediction for \pbGnineteenN{} second tiers,
written before their records were read, held (\pbGnineteenRho). They
ranked like their country's top league. Across rule changes in other
sports, leagues kept their order in \pbBeyondOrderHeldWord{} of
\pbBeyondOrderAskedWord{} tests. In every sport measured, a league
stayed near its place among its sport's leagues from season to season.
Under five, clubs inside a league differed more within a season than
the leagues did, but their differences lasted less. Rule-makers should judge
a change league by league, against each league's own seasons.
\end{abstract}


\noindent\textbf{Word count:} \mainWordCount{} words of main text, from
the Introduction to the Conclusion with headings, excluding tables,
references and figure captions.

\section{Introduction}

Every team sport has rules on who may play and when. Football caps its
substitutions, volleyball the players on its match sheet, handball its
team time-outs and cricket each bowler's overs.
Basketball and ice hockey let players come and go, and baseball now times
its pitchers. These rules change
often~\cite{arias2011_rule_modification,rennie2024_rule_changes_review}.
Yet studies of a change usually report one average effect on results,
running, injuries or the length of a game, for one league or several
combined~\cite{ayabe2022_j1_substitute_allowance,dilger2024_bundesliga_five_substitutions,zhang2025_substitution_rule_changes,gornall2026_in_game_rule_changes}.
Where one rule was studied in many countries, its effect differed between
them~\cite{moschini2010,dargahi2024}. So an average over leagues can hide
how each league responds. This paper asks what each league does with
the same rule, in seven sports.

A rule can say what teams may do, and we call this the room it gives
them. They choose what to do with that room. A speed limit is
the everyday example. It sets how fast drivers may go. Each road then
has its own usual speed below the limit. In the same way, each league
uses a rule's room in its own way. We call a league's own way of using
the rules its habit. Statisticians call it a standardised league effect
(Supplementary Section S6). A habit is measured from records the leagues
already keep, such as substitutions per match or the number of players a
team uses. The word describes only what teams do. A league's habit is the
sum of its coaches' decisions. If habits differ, and keep differing, a
rule-maker who reads the first season's average misjudges the leagues at
both ends.

Football offers the cleanest case. In the summer of 2020 dozens of
football leagues raised the substitution limit from three to five. IFAB,
football's rule-maker, had allowed five at FIFA's request, to protect
players' welfare during the
pandemic~\cite{ifab2020_five_temporary,ifab2022_five_permanent}. Each
competition chose whether and when to use
it~\cite{ifab2020_five_temporary}. So one new permission reached many
leagues at dates of their own. In 2022 IFAB made this permission
permanent~\cite{ifab2022_five_permanent,ifab2022_laws}. Every
substitution is recorded with its minute, so what each league did with
the new room can be counted.

Leagues are known to keep their own profiles, for example in home
advantage and in fouls and
cards~\cite{pollard2006_home_advantage,sapp2018_fouls_cards}. Analysts
have shown that the biggest football leagues kept the same order in
their use of substitutes from one season to the
next~\cite{caley2024_substitutes_leagues,segar2024_opta_substitute_trends}.
One of them called these differences lasting and explained them by
each league's customs~\cite{caley2024_substitutes_leagues}. This paper
tests whether leagues keep their order through a shared rule change, in
football and in six other sports. We wrote most of these tests down
before reading their data. We also wrote down \pbGdistinctWritten{}
predictions for seasons, leagues and competitions before reading their
records.

We tested four claims. First, under five substitutions football's
leagues would use the new room unevenly and keep their order. Second,
leagues in other sports would also keep their order in how they use
their rules. Third, a league's place among its sport's leagues would
persist from season to season about as strongly in every sport.
Fourth, a coach who moved would carry part of the old team's distance
from its league's average. If these held, a rule change would be best
judged league by league, against each league's own seasons.

\section{Methods}

\subsection{Sports, leagues and records}

Table~\ref{tab:sports} lists the sports in which these claims were
tested, the rule change studied in each, and their leagues and records.
Football's records come from
transfermarkt-datasets, a public collection built from Transfermarkt, a
football statistics website. We read its snapshot of
\pbSnapshotDate~\cite{cariboo2026_transfermarkt_datasets}. The
\pbFanLeagues{} European leagues whose
substitutions are counted hold \pbSeriesMatches{} matches from 2015--16
to 2025--26. Other leagues enter some measures (Table~\ref{tab:sports}).

Each league's limit in each season, three or five, was taken from the
rules in force and checked against the substitutions made (Supplementary
Table S1). Under five, a top-division team may substitute in at most
three breaks in play, called windows, and at
half-time~\cite{ifab2020_five_temporary,ifab2022_laws}. A season in which
the limit changed is split at the date of the change. We count
substitutions per team-match, which is one team's side of one match.
Supplementary Section S1 shows how the counts were checked against an
independent record, and corrected where the records fell short.

The records of basketball, volleyball, handball, ice hockey and baseball
come from the leagues' and federations' own sites. Cricket's T20 leagues
play matches of twenty overs a side. Their records come from Cricsheet,
a public archive of ball-by-ball records~\cite{cricsheet}.

\subsection{Ethics}

The study reused only public match records that already existed. These
records were its only source on players and managers, so informed
consent was outside its scope. Several tests read information related to
health: the injury reasons that the records give for some substitutions,
and players' absences from matchday squads. Both were counted by league,
and only league values are reported. The records name professional
players and managers, so they were treated as personal data, although
they are public. Processing was for scientific research under the
research provisions of the General Data Protection Regulation. Every
published table and figure is aggregated to leagues, or to summaries
across managers. The author treated these
published records as archival data, which \ethicsbody{} exempts from
approval~\cite{tcd2022_reams_guidance}. So the study ran under that
exemption. The same guidance grants approval only before a study begins.

\subsection{What is measured}

A rank correlation, or Spearman's, shows how well two orders agree: one
for the same order, zero for unrelated orders and minus one for the
reverse.

The spread between leagues in a season is the standard deviation of
their means: roughly, a typical league's gap from their mean. A league's
place on a measure is how many spreads its value sits above or below the
mean of its sport's leagues that season. The persistence $\phi$ is the
share of its place a league keeps from one season to the next. So after
$k$ seasons, a league keeps on average $\phi^k$ of its place. We call
this slow loss of place drift. We measured one $\phi$ for each sport. We used the correlations between
every pair of seasons, with the chance swings of single games taken out
(Supplementary Section S6).

A team's distance from its league is its substitutions per team-match
minus its league's mean that season. We average it over each coach's
spell. Basketball's distance uses the five busiest players' share of
minutes, and baseball's the players used beyond the starters. The coach's carry is the rank
correlation between the old team's distance and the new team's, across
the coaches who moved.

\subsection{Tests and statistics}

Each test was written down as a short text, most often a clause of its
own (Supplementary Section S3). The text fixes what the test measures,
its direction,
which leagues qualify and what each outcome would be called. Each was
saved, with its date, in the project's private online repository
(Supplementary Section S4).

The other sports were tested on a split record. Each game went to one of
two halves by a fixed rule that works like a coin toss. Half A was read
freely, to choose each measure and pass mark. The tests were then saved
before half B was read. The split guards against choosing a measure on
the luck of single games. But both halves share the same leagues, so a
measure chosen to suit those leagues could pass in both. The strongest
evidence is therefore the \pbGdistinctAsked{} predictions for new
seasons, leagues or competitions. Each was written down before any of its
records was read, and each was read once. One was rewritten before its
records were read, and it is counted once.

We call these two kinds of test pre-specified. Football's own tests,
with baseball's two on managers, were written during the analysis, after
their records had been read for other questions. Of the
\pbCClaimTests{} behind the claims, \pbCClaimBefore{} were saved
\pbCGapMinutes{} minutes to \pbCClaimHoursHigh{} hours before they were
run. The other \pbCClaimTogether{} were written down when they were run,
so they describe rather than test.

A 95\% interval gives the values that the data leave plausible
(Supplementary Section S1). $P$ is the chance that luck alone gives a
result at least this strong. Most tests are one-sided, because each
fixed its direction in its clause. Such a test passes only on a link in its
own direction. Every analysis added later is labelled stated. It
came after the tests, so it describes rather than
tests~\cite{ditroilo2025_sample_size}.

Some values combine the leagues of several sports. They come from
random-effects models, which give a combined value and the spread of the
leagues around it~\cite{riley2011_prediction_intervals}. They were
fitted after the split's tests, so their values describe rather than
test.

\subsection{Sample size estimation and justification}

The sources set the sample, so its size followed from them rather than
from an estimate made in advance. Every league that the sources cover
deeply enough to carry the measures was read. In football,
\pbFrameCovered{} of the source's \pbFrameLeagues{} top leagues have
seasons under three substitutions, so its coverage sets the sample
(Supplementary Section S1). That is a justification by the whole
available
population~\cite{lakens2022_sample_size,wagnsson2026_league_tables}. For
such a sample, this journal asks for the smallest effect each test could
detect~\cite{abt2025_sample_size}. The unit of each order test is the
league, so its sample size is its number of leagues
(Table~\ref{tab:register}).

For an order, the smallest effect of interest is a rank correlation of
\pbSsInterestRho. Below it, the order in one season accounts for less
than a quarter of the differences in a later one.

To detect a link, a test needs a \pbSsPowerTarget{} chance of passing at
a one-sided $P$ below \pbSsAlpha. Twelve leagues detect only a true rank correlation of
\pbSsMdeTwelve{} or more. Fourteen leagues detect \pbSsMdeFourteen{} or
more. So the order and explanation tests could detect only links well
above the smallest of interest. The largest test of coaches who moved
detects \pbSsMdeClubMovers{} or more. These values come from exact counts
up to nine leagues and from simulation beyond (Supplementary Section S1).

\section{Results}

The tests open with a check. As expected, each rule moved what it was
written to move (Table~\ref{tab:register}). The NHL, North America's top
ice hockey league, brought in three-on-three overtime to settle more tied
games before a shootout~\cite{nhl2015_three_on_three}. Between December
2014 and 2016--17, \pbKappaOTLeagues{} European leagues brought it in too. In
them, three-on-three overtime multiplied the odds of settling a tied game
before a shootout by \pbKappaOT. FIBA, basketball's world body, moved its
three-point line out in 2010~\cite{fiba2008_line_decision}. The move cut
each league's share of shots from three by about \pbLineRelPooled{} per
cent of that share. Baseball's pitch clocks were written to speed up the
game~\cite{mlb2023_pitch_clock}. They shortened games in all
\pbBeyondClockLeagues{} leagues tested. The size of the line's and the
clocks' effects differed between leagues (Supplementary Section S5). As
Meyer and Klatt found, football's two extra substitutions moved playing
time from starters to substitutes~\cite{meyer2024_england_contrast}.

\subsection{Football's leagues kept their order through the change}

Under three substitutions football's leagues sat close to their limit
(Figure~\ref{fig:fan}). As earlier counts found, a team made about
\pbFanThreeMean{} substitutions a match, close to three in every
league~\cite{meyer2021_one_additional_substitution,wei2024_substitutions,buchheit2023_substitutions}.
The leagues differed in how often teams used all three, from
\pbAtLimitThreeLow{} per cent of team-matches in
\pbAtLimitThreeLowLeague{} to \pbAtLimitThreeHigh{} per cent in
\pbAtLimitThreeHighLeague. So the leagues were already in an order, as
one analyst had seen in the biggest
leagues~\cite{caley2024_substitutes_leagues}.

That order carried through the change. The \pbPreOrderN{} leagues' order
over 2015--16 to 2018--19 was much the same as in their first season
under five (rank correlation \pbPreOrder). Under three, one season's
order predicted the next season's at \pbThreeOrderLow{} to
\pbThreeOrderHigh{} (Supplementary Section S2). So the order held about
as firmly as from one ordinary season to the next. These values were
read after the change, so they describe rather than test.

A prediction then tested the order in \pbGnineteenN{} European second
tiers. It was written before any of their records was collected. Their
order under three predicted their order under five (rank correlation
\pbGnineteenRho, one-sided $P = \pbGnineteenP$). In the seasons under
five they also ranked much as their own country's top league did. That
held for the \pbGnineteenLinkedN{} whose top leagues we had read (rank
correlation \pbGnineteenLinkedRho, one-sided $P = \pbGnineteenLinkedP$),
and for all \pbGnineteenPairedN{} (\pbGnineteenPairedRho; Supplementary
Section S4). So the order is partly a country's, shared by its first and
second tiers.

Under five, the top leagues spread apart in their substitutions, as
expected when a limit that most teams reach is
raised~\cite{kleven2016_bunching}. Had each league kept its share of the
limit, the spread would have grown by five thirds. It grew more than
that from 2020--21 to 2023--24, and about that much in the last two
seasons (Figure~\ref{fig:fan}b; Supplementary Section S2).

The change spread the leagues apart, and they kept their order. In the main
order test, saved \pbOrderTestHours{} hours before its run, the order in
the first season under five predicted
the order in 2025--26 (rank correlation \pbRho, one-sided
$P = \pbRhoP$). On the \pbOrderUnreadN{} leagues still unread
when the test was written, the rank correlation was \pbOrderUnread{}
($P = \pbOrderUnreadP$). The order from the seasons under three faded
slowly. By 2025--26 its rank correlation with the new order was
\pbOrdLatestFourteen, as slow drift would give (Supplementary
Section S2).

We tested four explanations of the order: money, rotation of the
starting eleven, the value of the bench and absences from matchday
squads. With thirteen or fourteen leagues, these tests could find only a strong link,
and every link was weaker than that. Inside leagues, richer clubs used
about as much of the limit as other clubs (mean rank correlation
\pbWealthMeanSigned, 95\% interval \pbWealthMeanSignedLow{} to
\pbWealthMeanSignedHigh). In all, \pbCandidatesAll{} tests read these
and other explanations. After a correction for the number of tests, only
\pbCandidatesSurvive{} passed: how often teams had used all three
windows with substitutions left. It largely restates how many
substitutions a league makes (Supplementary Section S3).

\subsection{Leagues in other sports kept their order in \pbBeyondOrderHeldWord{} of \pbBeyondOrderAskedWord{} tests}

Across rule changes in other sports, the leagues kept their order in
\pbBeyondOrderHeldWord{} of the \pbBeyondOrderAskedWord{} pre-specified
tests of the split record. Each test read few leagues. So only about
\pbSplitOrderExpected{} of the \pbBeyondOrderAskedWord{} would pass if
every order held at a rank correlation of \pbSsAtRho{}
(Table~\ref{tab:register}). Across FIBA's longer three-point line of
2010, basketball's leagues kept their order in the share of shots from
three. Across FIBA's shorter shot-clock reset of 2014, they kept it in
the five busiest players' share of minutes and in the number of players
given ten minutes. Volleyball's leagues kept theirs in the number of
players used.

The other tests failed. Basketball's pace, a team's possessions per
forty minutes, kept a weaker order. The 2014 change drew the leagues
together in pace, as another pre-specified test expected. Volleyball's
substitutions a set lost their order. Ice hockey's and baseball's tests
read four leagues each, where only a perfect order can pass. Ice
hockey's test read when a trailing team pulls its goalie, a choice its
rules leave to the coach (Supplementary Section S4).

Predictions also read orders across seasons in these sports, and
\pbGotherOrderHeld{} of \pbGotherOrderAsked{} held. Baseball's \pbGbaseballN{}
leagues kept their ranks in pitchers used per team and game. The leagues
lay far apart in pitchers used, so their prediction had expected this
with near certainty. Cricket's \pbGcricketN{} T20 leagues kept their
order in how many bowlers a side uses. The limit on each bowler's overs
stayed the same, and some leagues changed their substitute rules
(Tables~\ref{tab:sports} and~\ref{tab:register}). In basketball,
\pbGbasketOrderHeld{} of \pbGbasketOrderAsked{} such predictions held, and ice hockey's just
failed (Supplementary Section S6).

\subsection{A league's place persists in every sport measured}

As expected, in every sport measured, a league kept most of its place
among its sport's leagues from one season to the next
(Figure~\ref{fig:persistence}). Goals, points, volleyball's sets and
cricket's runs kept \pbBenchGoalsPhi{} to \pbBenchHockeyGoalsPhi{} of
their place a season, about as much. So leagues keep their character in
general~\cite{pollard2006_home_advantage,fort2007_structural_change},
and their habit is part of it. The persistence of basketball,
volleyball, ice hockey and baseball is consistent with one value,
\pbPhiPooled{} a season. That value rests mostly on basketball's leagues
(Figure~\ref{fig:persistence}a). Later seasons and new leagues, among
them the NBA and the WNBA, gave values close to it, as their predictions
expected (Table~\ref{tab:register}; Supplementary Section S6).

\subsection{Football's clubs differed more than its leagues within a season}

Under five, clubs inside a league differed more within one season than
the leagues did, as earlier charts
show~\cite{buchheit2023_substitutions,segar2024_opta_substitute_trends}.
Their spread was about \pbMagClubNoiseMatch{}
substitutions a match, against \pbSeasonSdFiveLow{} to
\pbSeasonSdFiveHigh{} between leagues. But a club kept about
\pbClubKeepOne{} of its distance from its league into the next season,
and a league about \pbLeagueKeepOne{} of its place. These values allow
for the chance swings of single matches (Supplementary Section S5).

\subsection{Coaches carried part of a team's distance from its league}

Coaches who moved carried part of their old team's distance from its
league (Figure~\ref{fig:carry}), as studies of managers' styles have
found
before~\cite{bertrand2003_managing,muehlheusser2018_managers,berry2021_leader_effects}.
Football managers carried their distance when they moved between clubs
of one league, and when they changed league. Those tests were written
down when they were run, so their values describe rather than test.
Basketball's coaches carried it too, on half B of the split record, and
a prediction on their moves since 2019 held (Figure~\ref{fig:carry}).
Baseball's \pbMoversBaseballTests{} tests, saved before they were run,
failed on its few moves, so baseball is inconclusive (Supplementary
Section S5).

The carry follows the coach rather than the club's choice of coach. In
football, a coach's distance at the old club correlated only
\pbSortingFootball{} with the new club's own distance in the season
before. Allowing for that earlier distance, the carry was
\pbCarryDestFootball{} in football and \pbCarryDestBasket{} in
basketball. A model that separates managers from their clubs found a
part that is the coach's own. It was small: about \pbMagCoachPerMatch{}
substitutions a match, or \pbMagCoachPerSeason{} a season (Supplementary
Section S5). These later checks are stated.

\section{Discussion}

Football's leagues kept their order through the change of limit, about
as firmly as from one ordinary season to the next. The \pbGnineteenN{}
second tiers read later kept theirs too. We read \pbGdistinctAsked{} of the \pbGdistinctWritten{}
predictions written before their records were read, and \pbGdistinctHeld{} held.
Of these passes, \pbPassesInformative{} could plausibly have failed,
among them the second tiers'. The other \pbPassesWeak{} would have passed
under a much weaker claim too (Table~\ref{tab:register}; Supplementary
Section S4).

These results fit how sport science sees a rule. A rule that gives room
without saying how to use it is a task
constraint~\cite{newell1986_constraints}. Loosening a constraint widens
the choices that work, and players and coaches choose among
them~\cite{renshaw2019_constraints_led,balague2019_constraints,ashford2020_rule_modifications}.
In this view a competition organises itself~\cite{mcgarry2002_dynamical}.
Football shows this at the scale of whole leagues. The old limit of
three held the leagues close to it. When teams are held at a limit, the count shows
only that they wanted at least that many~\cite{kleven2016_bunching}. So
raising the limit had to let the leagues spread apart. That part of the
spread is the limit's doing. But the order beneath the limit was already
there, and the limit did not set it.

Two readings fit the kept order. In one, each league raised its old
level in proportion to the new limit. In the other, each league answered
the change in its own way. The old limit held most teams at three, so
the leagues' old levels understate how differently they wanted to use
substitutes. Each league's level also drifts slowly, even when no rule
changes. So its departure from the leagues' shared rise lasts from
season to season (Supplementary Section S2). These counts therefore cannot tell
the two readings apart. The claim here is the one both share: the
leagues kept their order through the change.

Although clubs differ more within a season, a change is best judged by
league. A rule is adopted competition by competition. The Premier
League's clubs voted three times in 2020 against five substitutions,
partly from concern that five would favour clubs with deeper
squads~\cite{bbc2020_five_subs_vote,football365_2020_vote}. So the
league is where a change is adopted, and where differences in its use
last longer.

\subsection{Practical applications}

A rule-maker weighing a change, from IFAB to a single competition, can
expect the leagues that use the old rule most to use the new room most.
The old order matched the first season under five at a rank correlation
of \pbPreOrder, and 2025--26 at \pbOrdLatestFourteen. The overall rise
was known only after each season (Supplementary Section S2). The gaps
between the leagues are modest. In each season under
five, the spread between football's leagues was \pbSeasonSdFiveLow{} to
\pbSeasonSdFiveHigh{} substitutions per team-match. Over a season of
about \pbMagMatchesSeason{} league matches, that is about
\pbMagSeasonSubsLow{} to \pbMagSeasonSubsHigh{} substitutions per club.
Clubs within one league differed more, by about \pbMagClubNoiseSeason{}
substitutions a season, and a coach's own part was about
\pbMagCoachPerSeason{} (Supplementary Section S5). A new limit holds
teams back first where use is already near it. In 2025--26 the busiest
league used \pbTopShareFive{} per cent of five, and the least busy
\pbLowShareFive{} per cent.

So a change is best judged league by league. Methods for rules adopted
at different times estimate an effect for each group of leagues that
adopted it together~\cite{callaway2021_did}. Before the
change, measure how each league uses the rule, from the records it
already keeps. After it, compare each league with its own seasons
before, not with the average of all leagues. Judge it over several
seasons. After this change, use kept rising: every league made more
substitutions in 2025--26 than in its first season under five. Most
made about \pbUptakeMedian{} per cent of their rise to 2025--26 in the
first season (Supplementary Section S2). As reviews of rule changes advise, evaluate a change in
several competitions~\cite{gornall2026_in_game_rule_changes}, allow for
their differences and register the plan publicly before the records
arrive~\cite{rennie2024_rule_changes_review}. A study that combines
leagues to estimate a rule's effect on injuries or running should give
each league its own rise in substitutions, season by season. It should
also report the effect league by league.

\subsection{Relation to other work}

The closest report charted substitutions in seven of these leagues,
from the same source, up to 2021--22~\cite{buchheit2023_substitutions}.
It found them close to three a match under the old limit, and at four
to five under the new, apart from England, still under three. This
paper finds them already in an order, and asks whether it lasted.
Another study uses the same change to study coaches' decisions within
matches~\cite{cermak2026_negative}.

\ifanonymous
Several of these results first appeared in a preprint on nine
sports~\cite{ricou2026_one_equation}. That preprint also claimed a law
linking a league's load management to its timing. This paper drops it:
the two measures agree because of how they are built (Supplementary Section S7).
\else
Several of these results first appeared in the first version of a
preprint on nine sports~\cite{ricou2026_one_equation}. This paper is
that preprint's second version. The first version also claimed a law
linking a league's load management to its timing. This version drops it:
the two measures agree because of how they are built (Supplementary Section S7).
\fi

\subsection{Limitations}

No test here shows cause and effect. Every quantity describes how
playing time was recorded, and a coach's reasons enter only where a team
sheet records them. The records show how teams used their substitutions,
not how often players were injured. No habit here is linked to results,
running or injury. A test of the five-substitute rule's aim needs each
league's injuries and hours of play, recorded under football's consensus
methods~\cite{walden2023_football_consensus}.

The football records come from one public source, checked against an independent record only for some
leagues and seasons (Supplementary Section S1). In 2020--21 and 2021--22
many records were incomplete. The rule change also came with a
pandemic. But England, which kept three substitutions through the
crowded season of 2020--21, did not use more of them (Supplementary
Section S2).

Each league also sets rules of its own that were not measured in the
other sports. Examples are roster sizes, limits on foreign players and
MLB's limits on how many pitchers a team carries and how soon it may
change them~\cite{cbs2020_three_batter,ap2022_thirteen_pitchers}. These
alone could keep a league in its place.

Averaged over all team-matches, the windows held back at most
\pbHeldBackLow{} to \pbHeldBackHigh{} substitutions a match, by league.
So part of a league's shortfall from five may be the windows. Had every
team they stopped made all five, the order would barely have changed
(rank correlation \pbWindowOrderRho). The order was there under three,
before the windows (Supplementary Section S3).

The tests are small. Most rest on four to \pbIncrementN{} leagues, so
they could detect only a strong link. The predictions on persistence
would have failed only at a much weaker persistence, so their passes say
little (Supplementary Section S6).

No public register holds the tests. Their dates rest on
the project's private repository, whose history was rewritten
(Supplementary Section S4). For the tests written during the analysis,
a date shows when a test was saved, not when its records were first
read~\cite{vandenakker2021_secondary}. The split record's tests were uploaded with
their results, so only the order in which they were saved dates them.

When football's main order test was written down, the seasons under
five of France, Germany, Italy and Spain had already been read. On the
other \pbOrderUnreadN{} leagues alone it fell short of its pass mark. It
also measures persistence after the change, not across it (Supplementary
Section S2). The second tiers share their country's order, so they are
not an independent sample of leagues. In \pbTierSameBodyN{} of the
\pbGnineteenLinkedN{} countries one decision set both tiers' limit, so
what carries the order is open (Supplementary Section S4).

Two predictions for a football rule of 2026--27 were not read, because
this study ended before the end of that season (Table~\ref{tab:register};
Supplementary Section S4). The coach's carry rests on managers who
moved, and estimates from few movers can be
distorted~\cite{andrews2008_limited_mobility}. Supplementary Section S5
discusses the two published criticisms of the
design~\cite{fee2013_managers,jarosiewicz2023_revisiting}.

\section{Conclusion}

A speed limit sets how fast drivers may go, and each road has its own
usual speed below it. Football's substitution limit worked in a similar
way. When it rose, every league climbed. Most leagues made most of their
climb in the first season. At first they kept the order they already
had, and that order faded slowly, as drift would give. A prediction written before
\pbGnineteenN{} second tiers were read found that they kept their order
too. In every sport measured, a league's place persisted from season to
season. A rule sets the room, and each league uses
that room in its own way. So a rule change should be judged league by
league, against each league's own seasons before it, and over several
seasons.

\section*{Data availability}

All result tables, the analysis scripts and the test suite that rebuilds
every number in this paper are in the archived record at DOI
\archivedoi. The match records come from
third-party sources whose terms permit research use and not
redistribution. The result tables are the minimum dataset needed to
verify every claim. None of them names a player.

\section*{Disclosure of interest}

The author reports there are no competing interests to declare.

\section*{Funding}

This work received no specific funding.

\section*{Declaration of generative AI use}

Anthropic's Claude, a generative AI assistant, was used under the author's
direction to write and check the analysis scripts, search the literature
and edit the text. The version used for this revision was Claude Opus
5.5, through the Claude Code tool. It was used because a single author
could not otherwise build and check analyses of this size. The study
design, the claims and their
interpretation are the author's. Every reported quantity is computed from
the committed data by the archived scripts and was verified by the author,
who takes full responsibility for the content.

\bibliographystyle{vancouver_doi}
\bibliography{\referencelist}

\newcommand{\captionfan}{The \pbFanLeagues{} football leagues carried the order they had under
three substitutions through the change to five. (a) Substitutions per team-match in the
\pbFanLeagues{} European leagues, season by season, under the limit each
season's regulations set. Each league is named by its country. The three
highest leagues and the three lowest in 2025--26 are labelled, in the
darker greys. The other eight are in
light grey. England, dashed, used five in the restart of 2019--20 and
went back to three for 2020--21 and 2021--22. The band marks 2019--20,
which \pbFanMixedN{} leagues split at the date of their change.
Belgium's 2020--21 and Scotland's 2021--22 are split the same way, with
a small grey box behind each part. Each part of a split season is drawn
as its own point. Each mean is uncertain by about half a marker's
height under three, and by about a marker's height under five (one
standard error). Open markers are the \pbOverTenN{} league-seasons whose
records were incomplete for more than a tenth of their team-matches
(Supplementary Section S1). Several overlap in 2021--22. Ukraine's 2021--22 is not drawn, because almost none of its
team-matches were kept. (b) The same counts as a share of each season's
limit. On this scale, a league that kept its share of the limit stays
level. (c) Each league's mean under three, over 2015--16 to
2018--19, against its mean under five, over every season under five.
Each league keeps its marker from (a), and the leagues named in (a) are
named here. The rank correlation $\rho$ between the two means is printed
with its 95\% interval. It is description, read after the change. Open
points are Denmark and England, which are outside the order test
(Supplementary Section S2). The spread between the leagues depends on the
scale. When the leagues share a limit, their order is the same on every
scale (Supplementary Section S2).}
\newcommand{\captionpersistence}{In every sport measured, a league's place persists, about as firmly as
its scoring. (a) The persistence
$\phi$, the share of its place a league keeps from one season to the
next, with its 95\% interval. Beside it are the leagues and pairs of
seasons behind it, and each sport's weight in the pooled value, which
combines the sports. Squares are sports, with their measures together,
and circles are single measures. The open circle counts Japan's
volleyball sets as played (Supplementary Section S4). The diamond is the
pooled value over basketball, volleyball, ice hockey and baseball. Open
squares are predictions, written down before their records were read,
and they are not in the pool. Grey
points are scoring, for comparison: goals, points and cricket's runs. The dashed line is the pooled value, and the
dotted line a persistence of one. An interval that reaches past one
comes from few seasons. The scale at the top gives the half-life under
one rate: the number of seasons until a league would lose half of its
place. (b) Basketball, which carries most of the pooled weight. Each
point is the mean correlation of the leagues' places $k$ seasons apart,
with the chance swings of single games left in. Neighbouring values of
$k$ are grouped, and each point has its 95\% interval over leagues. Each
line is the mean correlation that a model predicts for the same pairs,
with those swings left in. The solid line has one rate, basketball's
$\phi$. The dashed line has a permanent part, about \pbTseShare{} per
cent of a league's place, plus a part that decays. It is fitted to the
leagues' seasons, not to these means. It runs a little above the points,
but it follows their levelling off, which one rate cannot.}
\newcommand{\captioncarry}{Coaches carried part of a team's distance from
its league. Each
row gives the rank correlation $\rho$ between two distances from the
league, with its 95\% interval. One is the distance of the team a coach
or manager left, and the other is the new team's. Beside it are the
slope of a straight-line fit, with its interval, and the number of
moves. Each sport has its own measure: substitutions a match
in football, the five busiest players' share of minutes in basketball,
and the players used beyond the starters in baseball. Football's
readings come from tests written down when they were run, so they
describe. Basketball's half of the games is the split record's test,
and its moves since 2019 a prediction. MLB's reading is a test saved
before its run. Black points,
filled or open, enter the pooled value, which combines the sports, with
their weight at right. The open point, the managers of MLB, the top
baseball league of the United States and Canada, is inconclusive. Grey
points are other readings of the same carry (Supplementary Section S5).
They are stated. The rows marked ``given the destination'' allow for the new club's own
distance in the season before. Basketball's readings of every game and
of half the games share most of their moves. The black diamond is the
pooled value over the three sports, with its 95\% interval, and the
dashed line marks it. Neither $\rho$ nor the slope is corrected for the
chance swings of single games. Those swings blur any link, so both
understate the carry.}

\clearpage
\begin{table}[p]
\centering
\caption{The seven sports, the rule change studied in each, and every record
the analyses read. Every league here enters at least one reported test or
estimate. The counts leave out the further football leagues that only
the explanation models read (Supplementary Section S1). The
\pbWomenPlacedN{} women's leagues of Supplementary Section S7 are left
out too, because no table counts their records.
Supplementary Tables~S1 and S2 give
each league's seasons. For football's first tiers, the number in
brackets gives the \pbFanLeagues{} European leagues whose substitutions
are counted. Football's other first tiers enter the rises over their own
seasons, or the timing measures (Supplementary Sections S2 and S7).
Seasons named by one year are played within a calendar year. Cricket's
bowling rule did not change. Supplementary Section S8 shows how each
competition shares out its minutes.}
\label{tab:sports}
\footnotesize
\input{tables/sports}
\end{table}

\clearpage
\begin{table}[p]
\centering
\turnsideways{\begin{minipage}{\sidewayswidth}
\captionsetup{font=scriptsize}
\caption{Every test of the paper's claims and checks,
failures included, in three blocks. The predictions on new records were
written down before any of their records was read. A prediction's
chance to pass, if its claim held, is the clause's own figure, marked
(clause). Otherwise it was computed after with the clause's simulator,
from the fits as they stood when the clause was written, marked (after).
Supplementary Section S4 gives each prediction's chance under a weaker
rival. One prediction of the 2010 line was written on the day the line's
first prediction failed, after that failure was recorded. Its spread was
fitted on the competitions read before, among them those the first
prediction read. The split record's tests were saved before half B was
read. These two blocks are pre-specified. The tests written during the
analysis came after their records had been read for other questions.
Saved before its run gives how many of a row's tests were saved before
they were run, \pbCGapMinutes{} minutes to \pbCGapHours{} hours before.
The others were written down when they were run, so they describe
rather than test. Supplementary Section S4 dates every test, and
Supplementary Section S7 describes the timing tests.
The letter after each block's name begins the labels of its tests in the
supplement.
The smallest attainable $P$ is the smallest $P$ a test can reach with
its number of leagues, or with the number of simulations used to
compute it. A dash marks a test with no such floor. Power is a rank test's
chance of passing when the true rank correlation is \pbSsAtRho, with a
one-sided $P$ below \pbSsAlpha{} to pass. $^{a}$The row includes a test
that could hardly pass, or hardly fail, with its leagues. $^{b}$Two of
these tests set the managers' share against labels shuffled across
team-matches, the wrong comparison (Supplementary Section S5). A third
reads mid-season changes, which follow poor results.}
\label{tab:register}
\registerfont
\input{tables/register_summary}
\end{minipage}}
\end{table}

\clearpage
\section*{Figure captions}

\begin{figure}[p]
\centering
\includegraphics[width=\textwidth]{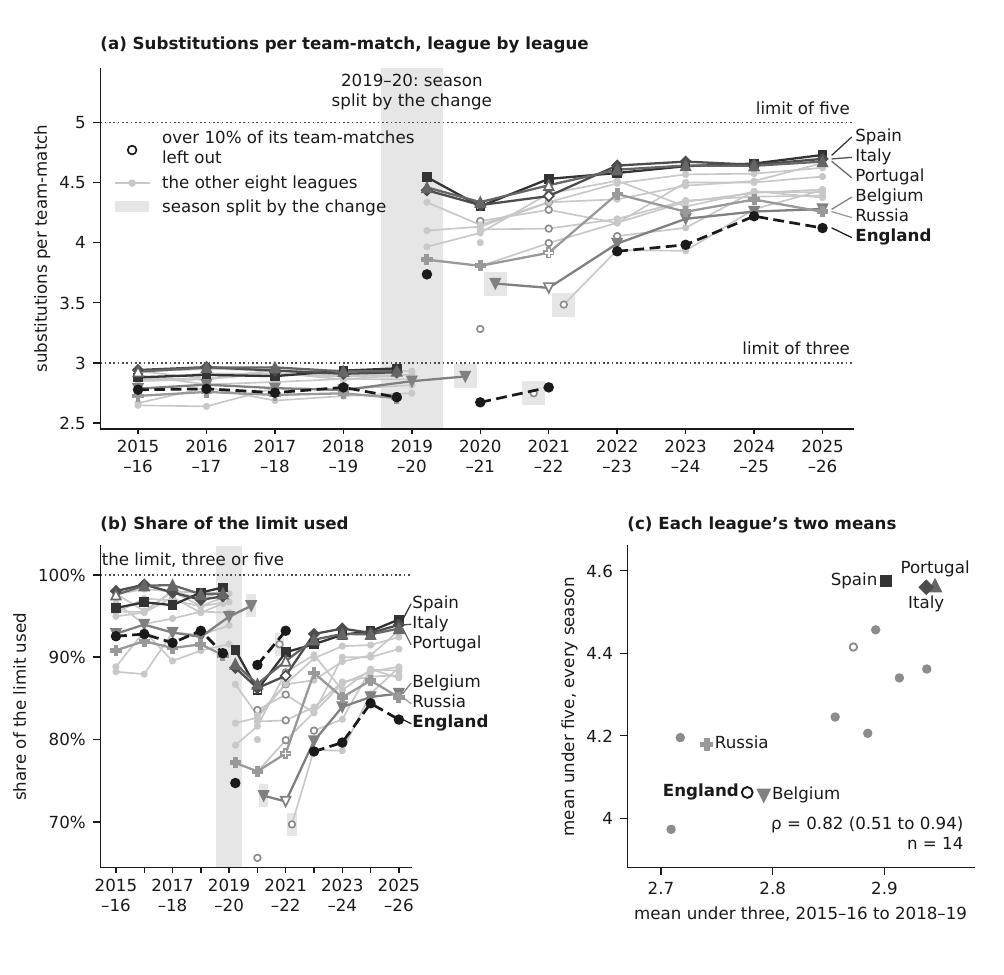}
\caption{}
\label{fig:fan}
\end{figure}

\noindent\textbf{Figure~\ref{fig:fan}.} \captionfan

\begin{figure}[p]
\centering
\includegraphics[width=\textwidth]{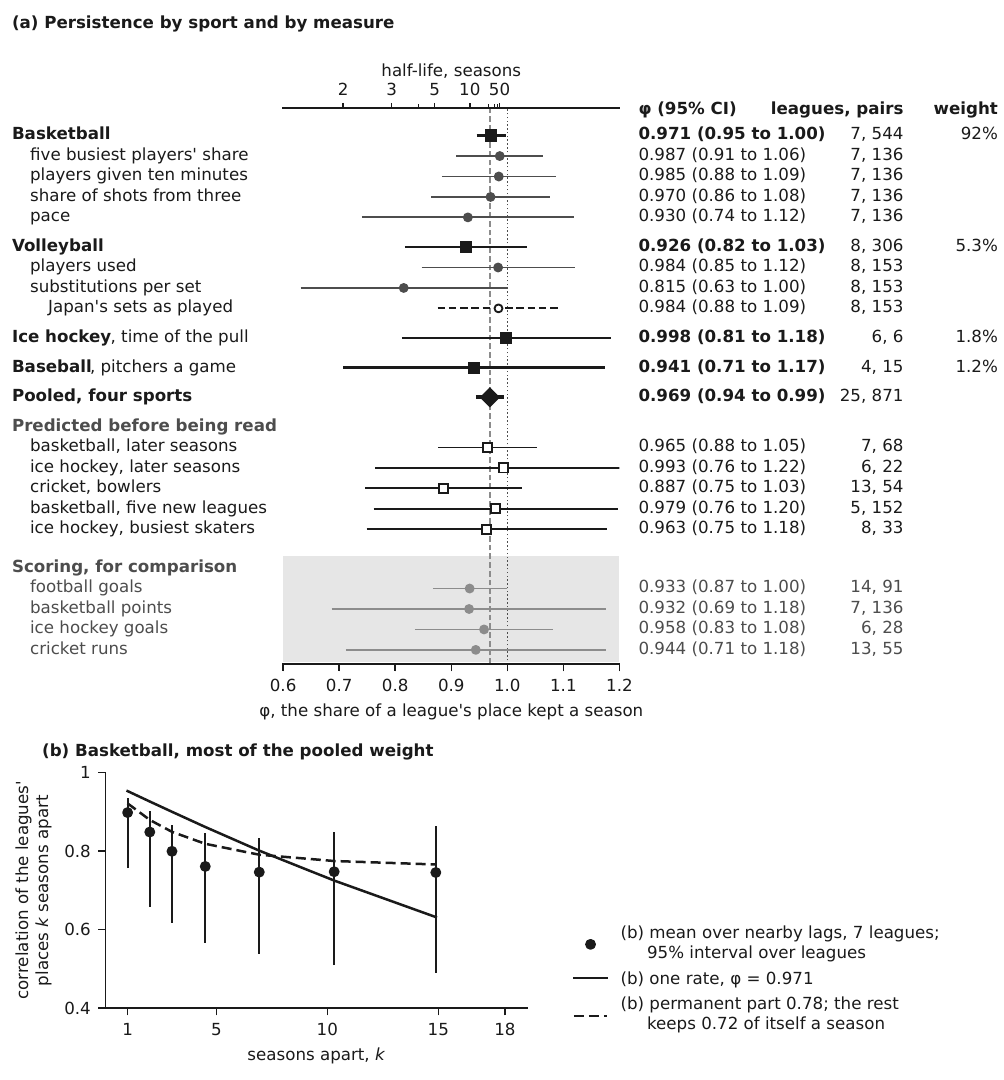}
\caption{}
\label{fig:persistence}
\end{figure}

\noindent\textbf{Figure~\ref{fig:persistence}.} \captionpersistence

\begin{figure}[p]
\centering
\includegraphics[width=\textwidth]{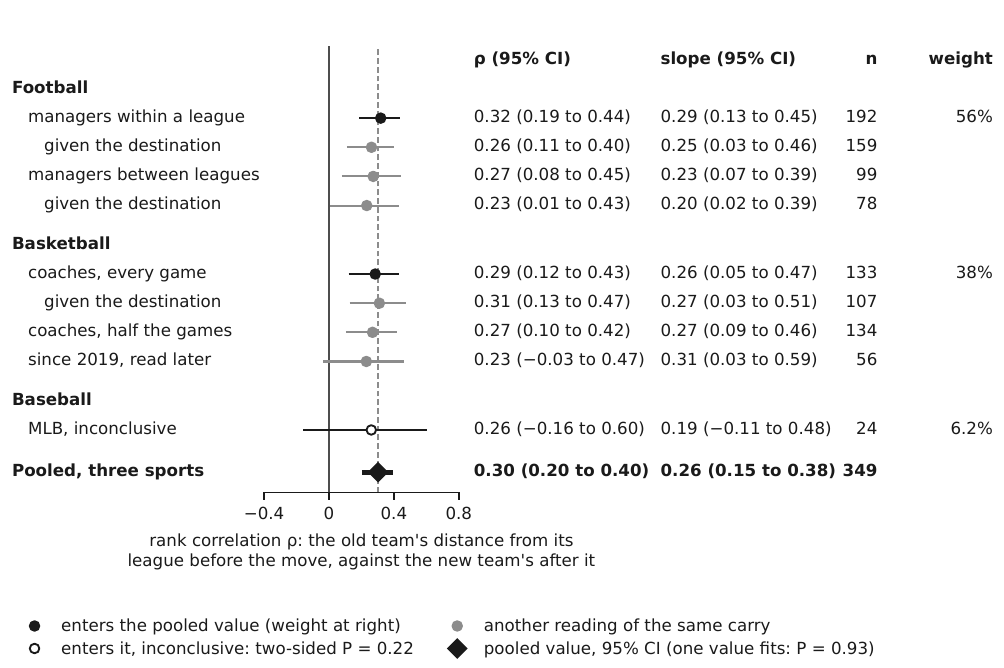}
\caption{}
\label{fig:carry}
\end{figure}

\noindent\textbf{Figure~\ref{fig:carry}.} \captioncarry

\clearpage

\clearpage

\clearpage

\end{document}

%% file: tables/numbers.tex
\newcommand{\pbRho}{0.73}%
\newcommand{\pbRhoP}{0.0033}%
\newcommand{\pbFanLeagues}{fourteen}%
\newcommand{\pbFanThreeMean}{2.85}%
\newcommand{\pbFanMixedN}{eight}%
\newcommand{\pbIncrementN}{17}%
\newcommand{\pbBeyondClockLeagues}{four}%
\newcommand{\pbKappaOT}{1.71}%
\newcommand{\pbKappaOTLeagues}{five}%
\newcommand{\pbPhiPooled}{0.969}%
\newcommand{\pbLineRelPooled}{9.9}%
\newcommand{\pbGbaseballN}{eight}%
\newcommand{\pbGcricketN}{thirteen}%
\newcommand{\pbSnapshotDate}{3 August 2026}%
\newcommand{\pbSeriesMatches}{43{,}159}%
\newcommand{\pbOverTenN}{fifteen}%
\newcommand{\pbPreOrder}{0.89}%
\newcommand{\pbPreOrderN}{fourteen}%
\newcommand{\pbThreeOrderLow}{0.80}%
\newcommand{\pbThreeOrderHigh}{0.87}%
\newcommand{\pbTseShare}{78}%
\newcommand{\pbBenchGoalsPhi}{0.93}%
\newcommand{\pbBenchHockeyGoalsPhi}{0.96}%
\newcommand{\pbWealthMeanSigned}{+0.02}%
\newcommand{\pbWealthMeanSignedLow}{\ensuremath{-}0.16}%
\newcommand{\pbWealthMeanSignedHigh}{+0.20}%
\newcommand{\pbCandidatesSurvive}{one}%
\newcommand{\pbCandidatesAll}{51}%
\newcommand{\pbCarryDestFootball}{0.26}%
\newcommand{\pbCarryDestBasket}{0.31}%
\newcommand{\pbSortingFootball}{0.08}%
\newcommand{\pbMagCoachPerMatch}{0.13}%
\newcommand{\pbMagCoachPerSeason}{4}%
\newcommand{\pbSsMdeTwelve}{0.69}%
\newcommand{\pbSsMdeFourteen}{0.65}%
\newcommand{\pbSsAlpha}{0.05}%
\newcommand{\pbSsPowerTarget}{0.80}%
\newcommand{\pbSsAtRho}{0.7}%
\newcommand{\pbSsMdeClubMovers}{0.20}%
\newcommand{\pbSsInterestRho}{0.5}%
\newcommand{\pbFrameLeagues}{thirty-one}%
\newcommand{\pbFrameCovered}{fourteen}%
\newcommand{\pbOrderUnread}{0.57}%
\newcommand{\pbOrderUnreadN}{eight}%
\newcommand{\pbOrderUnreadP}{0.076}%
\newcommand{\pbOrdLatestFourteen}{0.60}%
\newcommand{\pbSeasonSdFiveLow}{0.15}%
\newcommand{\pbSeasonSdFiveHigh}{0.30}%
\newcommand{\pbAtLimitThreeLow}{76}%
\newcommand{\pbAtLimitThreeHigh}{95}%
\newcommand{\pbAtLimitThreeLowLeague}{Scotland}%
\newcommand{\pbAtLimitThreeHighLeague}{Portugal}%
\newcommand{\pbHeldBackLow}{0.22}%
\newcommand{\pbHeldBackHigh}{0.42}%
\newcommand{\pbWindowOrderRho}{0.96}%
\newcommand{\pbGnineteenRho}{0.73}%
\newcommand{\pbGnineteenP}{0.0041}%
\newcommand{\pbGnineteenN}{twelve}%
\newcommand{\pbMagClubNoiseMatch}{0.36}%
\newcommand{\pbMagClubNoiseSeason}{12}%
\newcommand{\pbMagMatchesSeason}{33}%
\newcommand{\pbMagSeasonSubsLow}{5}%
\newcommand{\pbMagSeasonSubsHigh}{10}%
\newcommand{\pbTopShareFive}{94.6}%
\newcommand{\pbLowShareFive}{82.4}%
\newcommand{\pbUptakeMedian}{76}%
\newcommand{\pbGdistinctAsked}{seventeen}%
\newcommand{\pbGdistinctHeld}{twelve}%
\newcommand{\pbGdistinctWritten}{nineteen}%
\newcommand{\pbGotherOrderAsked}{seven}%
\newcommand{\pbGotherOrderHeld}{four}%
\newcommand{\pbGbasketOrderAsked}{four}%
\newcommand{\pbGbasketOrderHeld}{two}%
\newcommand{\pbMoversBaseballTests}{two}%
\newcommand{\pbPassesInformative}{seven}%
\newcommand{\pbPassesWeak}{five}%
\newcommand{\pbSplitOrderExpected}{four}%
\newcommand{\pbClubKeepOne}{0.52}%
\newcommand{\pbLeagueKeepOne}{0.91}%
\newcommand{\pbGnineteenLinkedN}{eight}%
\newcommand{\pbGnineteenLinkedRho}{0.90}%
\newcommand{\pbGnineteenLinkedP}{0.0023}%
\newcommand{\pbGnineteenPairedN}{twelve}%
\newcommand{\pbGnineteenPairedRho}{0.85}%
\newcommand{\pbTierSameBodyN}{five}%
\newcommand{\pbAllLeagues}{101}%
\newcommand{\pbWomenPlacedN}{two}%
\newcommand{\pbBeyondOrderHeldWord}{four}%
\newcommand{\pbBeyondOrderAskedWord}{eight}%
\newcommand{\pbCGapHours}{5}%
\newcommand{\pbCClaimTests}{20}%
\newcommand{\pbCClaimBefore}{seven}%
\newcommand{\pbCClaimTogether}{thirteen}%
\newcommand{\pbCClaimHoursHigh}{3}%
\newcommand{\pbCGapMinutes}{6}%
\newcommand{\pbOrderTestHours}{3}%

%% file: tables/counts.tex
\newcommand{\mainWordCount}{4,385}

%% file: tables/sports.tex
\begin{tabularx}{\textwidth}{>{\raggedright\arraybackslash}p{2.3cm}>{\raggedright\arraybackslash}X>{\raggedright\arraybackslash}p{2.6cm}rr}
\hline
Sport & Rule change read & Seasons & Leagues & Records \\
\hline
Football, first tiers & substitution limit raised from three to five, between 2019--20 and 2022--23 by league & 2015--16 to 2025--26 & 21 (14) & 47,937 matches \\
Football, second tiers & the same limit, in the second tiers of a prediction written before their records were read & 2015--16 to 2025--26 & 12 & 33,485 matches \\
Basketball & three-point line moved out, 2010--11; shot clock reset after an offensive rebound shortened, 2014--15 & 2006--07 to 2025 & 22 & 55,051 games \\
Volleyball & more players allowed on the match sheet, at dates that differ by league & 2006--07 to 2025--26 & 8 & 21,297 games \\
Handball & a third team time-out, an option each federation could take up; read in France, from January 2012 & 2011--12 to 2019--20 & 1 & 1,538 games \\
Ice hockey & tied games settled by three-on-three overtime before a shootout, at dates that differ by league & 2011--12 to 2025--26 & 12 & 44,312 games \\
Baseball & a pitch clock, enforced from 2023 to 2025 by league & 2008 to 2025 & 12 & 72,744 games \\
Cricket (T20) & none: at most a fifth of an innings per bowler throughout. Substitute rules changed in three leagues: the IPL's Impact Player from 2023, the ILT20's Super-Sub from 2023--24 and the BBL's X-Factor in 2020--21 and 2021--22 & 2019 to 2025 & 13 & 3,366 innings \\
\hline
Six other sports & & & 68 & 198,308 games and innings \\
Seven sports & & & 101 & 279,730 matches, games and innings \\
\hline
\end{tabularx}

%% file: tables/register_summary.tex
\setlength{\tabcolsep}{3pt}
\renewcommand{\arraystretch}{0.88}
\begin{tabularx}{\linewidth}{>{\raggedright\arraybackslash}X>{\raggedright\arraybackslash}p{3.1cm}c>{\raggedright\arraybackslash}p{2.8cm}>{\raggedright\arraybackslash}p{1.4cm}>{\raggedright\arraybackslash}p{2.1cm}>{\raggedright\arraybackslash}p{1.5cm}>{\raggedright\arraybackslash}p{1.9cm}}
\hline
Claim & Asked in & Leagues & Smallest attainable $P$ & Power at 0.7 & Chance to pass if the claim held & Saved before its run & Held / asked \\
\hline
\multicolumn{8}{l}{\textit{predictions on new records (G)}} \\
One constant per rule (the 2010 line; three-on-three overtime) & basketball; ice hockey & 1--2 & -- & -- & 0.73 to 0.91 (after) & all & 1 of 2 \\
One proportion of the share before, with a spread (the 2010 line) & basketball & 6 & -- & -- & 0.95 (after) & all & 1 of 1 \\
One constant per second of clock (the first pitch clocks) & baseball & 6 & -- & -- & 0.80 (clause) & all & 0 of 1 \\
A league keeps its place (basketball) & basketball & 7 & \ensuremath{2.0 \times 10^{-4}} & 0.49 & 0.48 to 0.50 (after) & all & 2 of 4 \\
The order holds over more leagues (ice hockey, baseball, cricket)$^{a}$ & ice hockey; baseball; cricket & 6--13 & \ensuremath{2.5 \times 10^{-5}} to 0.0014 & 0.32 to 0.85 & 0.39 to 1.00 (clause) & all & 2 of 3 \\
One persistence a season across sports$^{a}$ & ice hockey; basketball; cricket & 5--13 & -- & -- & 0.91 (after) to 1.00 (clause) & all & 4 of 4 \\
One carry across sports (the coach) & basketball & 3 & \ensuremath{1.0 \times 10^{-4}} & 1.00 & 0.69 (after) & all & 1 of 1 \\
Football's rules in new seasons and new leagues (written down for this version)$^{a}$ & football & 12--14 & \ensuremath{1.0 \times 10^{-5}} & -- & 0.88 to 0.97 (clause) & all & 1 of 1 (2 not read) \\
\multicolumn{8}{l}{\textit{split record (R)}} \\
A rule moves what it was written to move & basketball; ice hockey; baseball & 4--6 & -- & -- & -- & all & 3 of 3 \\
The order of the leagues persists$^{a}$ & basketball; volleyball; ice hockey; baseball & 4--8 & \ensuremath{2.5 \times 10^{-5}} to 0.042 & 0.22 to 0.60 & -- & all & 4 of 8 \\
A tightening draws the leagues together & basketball; baseball & 4--7 & \ensuremath{5.0 \times 10^{-4}} & -- & -- & all & 1 of 3 \\
A loosened limit is used more, and lets the leagues differ & handball; volleyball; basketball; ice hockey & 1--8 & \ensuremath{5.0 \times 10^{-4}} & -- & -- & all & 4 of 6 \\
A climb slows and levels off & handball; ice hockey; baseball & 1--4 & -- & -- & -- & all & 0 of 3 \\
The habit before a change predicts the response & volleyball & 5 & -- & -- & -- & all & 0 of 1 \\
The coach carries part of a team's distance from its league & basketball & 3 & \ensuremath{1.0 \times 10^{-4}} & 1.00 & -- & all & 2 of 2 \\
\multicolumn{8}{l}{\textit{tests written during the analysis (C)}} \\
The order persists across the five-substitution era (the order test) & football & 12 & \ensuremath{2.1 \times 10^{-9}} & 0.81 & -- & 1 of 1 & 1 of 1 \\
A material explanation accounts for the order (money, rotation, bench value, absences) & football & 12--14 & \ensuremath{4.8 \times 10^{-14}} to \ensuremath{1.6 \times 10^{-10}} & 0.85 to 0.92 & -- & 3 of 4 & 0 of 4 \\
A further explanation accounts for the order (culture; the league's own increment; two batteries) & football & 13--31 & \ensuremath{1.2 \times 10^{-34}} to \ensuremath{1.6 \times 10^{-10}} & 0.85 to 0.94 & -- & 1 of 8 & 0 of 8 \\
The coach's carry, football$^{b}$ & football & 14 & -- & 1.00 & -- & 0 of 5 & 5 of 5 \\
The coach's carry, baseball (MLB's managers) & baseball & 1 & \ensuremath{1.6 \times 10^{-24}} to \ensuremath{1.0 \times 10^{-3}} & 0.97 & -- & 2 of 2 & 0 of 2 \\
Timing before the change (the pre-change tests)$^{a}$ & football & 8--12 & \ensuremath{2.5 \times 10^{-5}} & 0.60 & -- & 4 of 4 & 1 of 4 \\
The timing slope on separate halves of the matches$^{a}$ & football & 12 & -- & -- & -- & 1 of 1 & 1 of 1 \\
Resting between matches against timing within them$^{a}$ & football & 8 & \ensuremath{2.5 \times 10^{-5}} & 0.60 & -- & 1 of 1 & 0 of 1 \\
\hline
\end{tabularx}

%% file: manuscript_b.bbl
\begin{thebibliography}{10}

\bibitem{arias2011_rule_modification}
Arias JL, Argudo FM, Alonso JI.
\newblock Review of rule modification in sport.
\newblock Journal of Sports Science and Medicine. 2011;10(1):1-8.

\bibitem{rennie2024_rule_changes_review}
Rennie G, Chesson L, Weaving D, Jones B.
\newblock The effects of rule changes in football-code team sports: a
  systematic review.
\newblock Science and Medicine in Football. 2025;9(3):199-212.
\newblock doi:\url{10.1080/24733938.2024.2375752}.

\bibitem{ayabe2022_j1_substitute_allowance}
Ayabe M, Sunami S, Kumahara H, Ishizaki S.
\newblock Effects of substitute allowance on match activity characteristics in
  {Japanese} professional football across 2019, 2020, and 2021 seasons.
\newblock Journal of Sports Sciences. 2022;40(23):2654-60.
\newblock doi:\url{10.1080/02640414.2023.2182878}.

\bibitem{dilger2024_bundesliga_five_substitutions}
Dilger A, Vischer L.
\newblock Effects of the rule change from three to five substitutions in the
  {Bundesliga}.
\newblock International Journal of Sport Finance. 2024;19(4):198-207.
\newblock doi:\url{10.32731/ijsf.194.112024.01}.

\bibitem{zhang2025_substitution_rule_changes}
Zhang H, Mathes J, Garnica~Caparr{\'o}s M, Memmert D.
\newblock Assessing the impact of substitution rule changes in {European}
  football leagues.
\newblock International Journal of Sports Science \& Coaching. 2025.
\newblock Published online 21 February 2025.
\newblock doi:\url{10.1177/17479541251316162}.

\bibitem{gornall2026_in_game_rule_changes}
Gornall H, Truscott H, Shill IJ, Ashford M, Palmer D.
\newblock Evaluating the use of in-game rule changes as a primary prevention
  approach to reduce injury risk in invasion team sports: a scoping review.
\newblock Sports Medicine. 2026;56(4):927-39.
\newblock doi:\url{10.1007/s40279-026-02405-8}.

\bibitem{moschini2010}
Moschini G.
\newblock Incentives and outcomes in a strategic setting: the
  3-points-for-a-win system in soccer.
\newblock Economic Inquiry. 2010;48(1):65-79.
\newblock doi:\url{10.1111/j.1465-7295.2008.00177.x}.

\bibitem{dargahi2024}
Dargahi S, Reilly B.
\newblock Assessing an experimental rule change in the pay-offs for soccer
  league match outcomes using historical data for {Ireland}.
\newblock Frontiers in Behavioral Economics. 2024;3.
\newblock doi:\url{10.3389/frbhe.2024.1506963}.

\bibitem{ifab2020_five_temporary}
{The International Football Association Board}. Five substitutes option
  temporarily allowed for competition organisers; 2020.
\newblock The IFAB, news, 8 May 2020.
\newblock Available from:
  \url{https://www.theifab.com/news/five-substitutes-option-temporarily-allowed-for-competition-organisers}.

\bibitem{ifab2022_five_permanent}
{The International Football Association Board}. The {IFAB} permanently approves
  five-substitute option in top-level competitions; 2022.
\newblock The IFAB, news, 13 June 2022.
\newblock Available from:
  \url{https://www.theifab.com/news/the-ifab-permanently-approves-five-substitute-option-in-top-level-competitions/}.

\bibitem{ifab2022_laws}
{The International Football Association Board}.
\newblock Laws of the Game 2022/23.
\newblock Zurich: The International Football Association Board; 2022.

\bibitem{pollard2006_home_advantage}
Pollard R.
\newblock Worldwide regional variations in home advantage in association
  football.
\newblock Journal of Sports Sciences. 2006;24(3):231-40.
\newblock doi:\url{10.1080/02640410500141836}.

\bibitem{sapp2018_fouls_cards}
Sapp RM, Spangenburg EE, Hagberg JM.
\newblock Trends in aggressive play and refereeing among the top five
  {European} soccer leagues.
\newblock Journal of Sports Sciences. 2018;36(12):1346-54.
\newblock doi:\url{10.1080/02640414.2017.1377911}.

\bibitem{caley2024_substitutes_leagues}
Caley M. Substitutes, {Part II}: clubs, leagues and managers; 2024.
\newblock Expecting Goals, 8 February 2024.
\newblock Available from:
  \url{https://www.expectinggoals.com/p/substitutes-part-ii-clubs-leagues}.

\bibitem{segar2024_opta_substitute_trends}
Segar D. How substitute trends compare across {Europe's} top five leagues;
  2024.
\newblock Opta Analyst, 28 March 2024.
\newblock Available from:
  \url{https://theanalyst.com/articles/how-substitute-trends-compare-across-europe-top-five-leagues}.

\bibitem{cariboo2026_transfermarkt_datasets}
Cariboo D. transfermarkt-datasets: clean, structured football data built from
  {Transfermarkt}; 2026.
\newblock GitHub repository, commit af8c44a, snapshot of 3 August 2026,
  CC0-1.0.
\newblock Available from:
  \url{https://github.com/dcaribou/transfermarkt-datasets}.

\bibitem{cricsheet}
{Cricsheet}. Ball-by-ball data for limited-overs cricket; 2026.
\newblock cricsheet.org, archive of all matches in JSON, downloaded 6 September
  2026.

\bibitem{tcd2022_reams_guidance}
{Trinity College Dublin}. Guidance for using the {Research} {Ethics}
  {Application} {Management} {System} ({REAMS}): making an application; 2022.
\newblock Institutional guidance, pages 5 and 27--28. Accessed 25 September
  2026.
\newblock Available from:
  \url{https://www.tcd.ie/research/support/assets/pdf/Top%20Guidance_Simplified-31st%20Oct.pdf}.

\bibitem{ditroilo2025_sample_size}
Ditroilo M, Mesquida C, Abt G, Lakens D.
\newblock Exploratory research in sport and exercise science: Perceptions,
  challenges, and recommendations.
\newblock Journal of Sports Sciences. 2025;43(12):1108-20.
\newblock doi:\url{10.1080/02640414.2025.2486871}.

\bibitem{riley2011_prediction_intervals}
Riley RD, Higgins JPT, Deeks JJ.
\newblock Interpretation of random effects meta-analyses.
\newblock BMJ. 2011;342:d549.
\newblock doi:\url{10.1136/bmj.d549}.

\bibitem{lakens2022_sample_size}
Lakens D.
\newblock Sample size justification.
\newblock Collabra: Psychology. 2022;8(1):33267.
\newblock doi:\url{10.1525/collabra.33267}.

\bibitem{wagnsson2026_league_tables}
Wagnsson S, Lundkvist E.
\newblock Scoreless but still scoring: A population-based observational study
  of removing league tables and playoffs in children's soccer.
\newblock Journal of Sports Sciences. 2026;44(19):2709-19.
\newblock doi:\url{10.1080/02640414.2026.2713886}.

\bibitem{abt2025_sample_size}
Abt G, Boreham C, Davison G, Jackson R, Jobson S, Wallace E, et~al.
\newblock Sample size estimation revisited.
\newblock Journal of Sports Sciences. 2025;43(21):2511-6.
\newblock doi:\url{10.1080/02640414.2025.2499403}.

\bibitem{nhl2015_three_on_three}
Rosen D. Board of {Governors} {OKs} 3-on-3 {OT}, coach's challenge; 2015.
\newblock NHL.com, 24 June 2015; Internet Archive copy of 31 October 2015
  (accessed 15 October 2026).
\newblock Available from:
  \url{https://web.archive.org/web/20151031130948/http://www.nhl.com/ice/news.htm?id=771982}.

\bibitem{fiba2008_line_decision}
{International Basketball Federation (FIBA)}. {FIBA} approves historic rule
  changes; 2008.
\newblock Press release, Geneva, 26 April 2008, as republished by Basketball
  Manitoba on 29 April 2008 (accessed 15 October 2026).
\newblock Available from:
  \url{https://www.basketballmanitoba.ca/2008/04/fiba-approves-historic-rule-changes.html}.

\bibitem{mlb2023_pitch_clock}
{Major League Baseball}. {MLB} announces rule changes for 2023 season; 2022.
\newblock Press release, 9 September 2022 (accessed 15 October 2026).
\newblock Available from:
  \url{https://www.mlb.com/press-release/press-release-mlb-announces-rule-changes-for-2023-season}.

\bibitem{meyer2024_england_contrast}
Meyer J, Klatt S.
\newblock Additional substitutions in elite {European} football.
\newblock International Journal of Sports Science \& Coaching.
  2024;19(2):769-78.
\newblock doi:\url{10.1177/17479541231164090}.

\bibitem{meyer2021_one_additional_substitution}
Meyer J, Klatt S.
\newblock Impact of one additional substitution on player load and coaching
  tactics in elite football.
\newblock Applied Sciences. 2021;11(16):7676.
\newblock doi:\url{10.3390/app11167676}.

\bibitem{wei2024_substitutions}
Wei X, Shu Y, Liu J, Chmura P, Randers MB, Krustrup P.
\newblock Analysing substitutions in recent {World} {Cups} and {European}
  {Championships} in male and female elite football -- influence of new
  substitution rules.
\newblock Biology of Sport. 2024;41(3):267-74.
\newblock doi:\url{10.5114/biolsport.2024.134755}.

\bibitem{buchheit2023_substitutions}
Buchheit M, Settembre M, Hader K, Tarascon A, McHugh D, Verheijen R.
  Know-your-own-League context: insights for player preparation and recruitment
  -- {Part 3}: Player substitutions; 2023.
\newblock Sport Performance \& Science Reports, no. 185, v1, March 2023.
\newblock Available from:
  \url{https://sportperfsci.com/wp-content/uploads/2023/03/SPSR185_Buchheit.pdf}.

\bibitem{kleven2016_bunching}
Kleven HJ.
\newblock Bunching.
\newblock Annual Review of Economics. 2016;8(1):435-64.
\newblock doi:\url{10.1146/annurev-economics-080315-015234}.

\bibitem{fort2007_structural_change}
Fort R, Lee YH.
\newblock Structural change, competitive balance, and the rest of the major
  leagues.
\newblock Economic Inquiry. 2007;45(3):519-32.
\newblock doi:\url{10.1111/j.1465-7295.2007.00026.x}.

\bibitem{bertrand2003_managing}
Bertrand M, Schoar A.
\newblock Managing with Style: The Effect of Managers on Firm Policies.
\newblock The Quarterly Journal of Economics. 2003;118(4):1169-208.
\newblock doi:\url{10.1162/003355303322552775}.

\bibitem{muehlheusser2018_managers}
Muehlheusser G, Schneemann S, Sliwka D, Wallmeier N.
\newblock The contribution of managers to organizational success: evidence from
  {German} soccer.
\newblock Journal of Sports Economics. 2018;19(6):786-819.
\newblock doi:\url{10.1177/1527002516674760}.

\bibitem{berry2021_leader_effects}
Berry CR, Fowler A.
\newblock Leadership or luck? {Randomization} inference for leader effects in
  politics, business, and sports.
\newblock Science Advances. 2021;7(4):eabe3404.
\newblock doi:\url{10.1126/sciadv.abe3404}.

\bibitem{newell1986_constraints}
Newell KM.
\newblock Constraints on the development of coordination.
\newblock In: Wade MG, Whiting HTA, editors. Motor Development in Children:
  Aspects of Coordination and Control. Dordrecht: Martinus Nijhoff; 1986. p.
  341-60.
\newblock doi:\url{10.1007/978-94-009-4460-2_19}.

\bibitem{renshaw2019_constraints_led}
Renshaw I, Davids K, Newcombe D, Roberts W.
\newblock The Constraints-Led Approach: Principles for Sports Coaching and
  Practice Design.
\newblock Abingdon: Routledge; 2019.
\newblock doi:\url{10.4324/9781315102351}.

\bibitem{balague2019_constraints}
Balagu{\'e} N, Pol R, Torrents C, Ric A, Hristovski R.
\newblock On the relatedness and nestedness of constraints.
\newblock Sports Medicine - Open. 2019;5(1):6.
\newblock doi:\url{10.1186/s40798-019-0178-z}.

\bibitem{ashford2020_rule_modifications}
Ashford M, Burke K, Barrell D, Abraham A, Poolton J.
\newblock The impact of rule modifications on player behaviour in a talent
  identification and development environment: A case study of the {Rugby
  Football Union's Wellington Academy Rugby Festival}.
\newblock Journal of Sports Sciences. 2020;38(23):2670-6.
\newblock doi:\url{10.1080/02640414.2020.1795559}.

\bibitem{mcgarry2002_dynamical}
McGarry T, Anderson DI, Wallace SA, Hughes MD, Franks IM.
\newblock Sport competition as a dynamical self-organizing system.
\newblock Journal of Sports Sciences. 2002;20(10):771-81.
\newblock doi:\url{10.1080/026404102320675620}.

\bibitem{bbc2020_five_subs_vote}
{BBC Sport}. Premier League clubs vote against use of five substitutes next
  season; 2020.
\newblock BBC Sport via TNT Sports, 6 August 2020.
\newblock Available from:
  \url{https://www.tntsports.co.uk/football/premier-league/2019-2020/premier-league-clubs-vote-against-use-of-5-substitutes-next-season-bbc_sto7828108/story.shtml}.

\bibitem{football365_2020_vote}
{Football365}. {PL} clubs vote against five subs; nine allowed on bench; 2020.
\newblock Football365, 17 December 2020.
\newblock Available from:
  \url{https://www.football365.com/news/premier-league-five-subs-vote-nine-bench}.

\bibitem{callaway2021_did}
Callaway B, Sant'Anna PHC.
\newblock Difference-in-differences with multiple time periods.
\newblock Journal of Econometrics. 2021;225(2):200-30.
\newblock doi:\url{10.1016/j.jeconom.2020.12.001}.

\bibitem{cermak2026_negative}
Cermak L.
\newblock The Negative Productivity of Strategic Flexibility: Evidence from a
  Natural Experiment in Football; 2026.
\newblock SSRN working paper.
\newblock doi:\url{10.2139/ssrn.6983270}.

\bibitem{ricou2026_one_equation}
Ricou GP. One equation allocates playing time in competitive team sports: the
  rulebook sets the arithmetic and the league sets the habit; 2026.
\newblock arXiv:2609.05584v1 [physics.soc-ph].

\bibitem{walden2023_football_consensus}
Wald{\'e}n M, Mountjoy M, McCall A, Serner A, Massey A, Tol JL, et~al.
\newblock Football-specific extension of the {IOC} consensus statement: methods
  for recording and reporting of epidemiological data on injury and illness in
  sport 2020.
\newblock British Journal of Sports Medicine. 2023;57(21):1341-50.
\newblock doi:\url{10.1136/bjsports-2022-106405}.

\bibitem{cbs2020_three_batter}
{CBS Sports}. {MLB} announces rule changes for 2020 season, including
  three-batter minimum and two-way player designation; 2020.
\newblock CBS Sports, 13 February 2020.
\newblock Available from:
  \url{https://www.cbssports.com/mlb/news/mlb-announces-rule-changes-for-2020-season-including-three-batter-minimum-and-two-way-player-designation/}.

\bibitem{ap2022_thirteen_pitchers}
Cohen J. {MLB} moving forward with long-delayed 13-pitcher limit; 2022.
\newblock Associated Press, via KSL.com, 19 June 2022.
\newblock Available from:
  \url{https://www.ksl.com/article/50425882/mlb-moving-forward-with-long-delayed-13-pitcher-limit}.

\bibitem{vandenakker2021_secondary}
van~den Akker OR, Weston S, Campbell L, Chopik B, Damian R, Davis-Kean P,
  et~al.
\newblock Preregistration of secondary data analysis: a template and tutorial.
\newblock Meta-Psychology. 2021;5.
\newblock doi:\url{10.15626/MP.2020.2625}.

\bibitem{andrews2008_limited_mobility}
Andrews MJ, Gill L, Schank T, Upward R.
\newblock High wage workers and low wage firms: negative assortative matching
  or limited mobility bias?
\newblock Journal of the Royal Statistical Society: Series A (Statistics in
  Society). 2008;171(3):673-97.
\newblock doi:\url{10.1111/j.1467-985X.2007.00533.x}.

\bibitem{fee2013_managers}
Fee CE, Hadlock CJ, Pierce JR.
\newblock Managers with and without Style: Evidence Using Exogenous Variation.
\newblock The Review of Financial Studies. 2013;26(3):567-601.
\newblock doi:\url{10.1093/rfs/hhs131}.

\bibitem{jarosiewicz2023_revisiting}
Jarosiewicz VE, Ross DG.
\newblock Revisiting Managerial ``Style'': The Replicability and Falsifiability
  of Manager Fixed Effects for Firm Policies.
\newblock Strategic Management Journal. 2023;44(3):858-86.
\newblock doi:\url{10.1002/smj.3439}.

\end{thebibliography}
